**Can a Natural Image-Based Foundation Model Outperform a Retina-Specific Model in Detecting Ocular and Systemic Diseases?**

Qingshan Hou[1,2,3*], Yukun Zhou[4,5,6*], Jocelyn Hui Lin Goh[1,3,7*], Ke Zou[1,3], Samantha Min Er Yew[1,3], Sahana Srinivasan[1,3], Meng Wang[1,3], Thaddaeus Lo[7], Xiaofeng Lei[8], Siegfried K. Wagner[5,9], Mark A. Chia[5,9], Gabriel Dawei Yang[7], Hongyang Jiang[10], An Ran Ran[10], Rui Santos[11,12], Gabor Mark Somfai[11,12,13], Juan Helen Zhou[14,15], Haoyu Chen[16,17], Qingyu Chen[18], Carol Y. Cheung[10], Pearse A. Keane[5,9]†, Yih Chung Tham[1,3,7,19]†

***contributed equally as first author**

**†contributed equally as last author**

[1] Department of Ophthalmology, Yong Loo Lin School of Medicine, National University of Singapore, Singapore

[2] School of Computer Science and Engineering, Northeastern University, Shenyang, China

[3] Centre for Innovation and Precision Eye Health, Yong Loo Lin School of Medicine, National University of Singapore, Singapore

[4] Centre for Medical Image Computing, University College London, London, United Kingdom.

[5] NIHR Biomedical Research Centre at Moorfields Eye Hospital NHS Foundation Trust, London, UK

[6]Department of Medical Physics and Biomedical Engineering, University College London, London, United Kingdom.

[7] Singapore Eye Research Institute, Singapore National Eye Centre, Singapore

[8] Institute of High-Performance Computing (IHPC), Agency for Science, Technology and Research (A*STAR), Singapore, Singapore.

[9] Institute of Ophthalmology, University College London, London, UK

[10] Department of Ophthalmology and Visual Sciences, The Chinese University of Hong Kong, Hong Kong, China

[11] Department of Ophthalmology, Stadtspital Zürich, Zurich, Switzerland

[12] Spross Research Institute, Zurich, Switzerland

[13] Department of Ophthalmology, Semmelweis University, Budapest, Hungary

[14] Centre for Sleep and Cognition & Centre for Translational MR Research, Yong Loo Lin School of Medicine, National University of Singapore, Singapore

[15] Department of Electrical and Computer Engineering, National University of Singapore, Singapore

[16] Joint Shantou International Eye Center, Shantou University and the Chinese University of Hong Kong, Shantou, Guangdong, China

[17] Shantou University Medical College, Shantou, Guangdong, China

[18] Department of Biomedical Informatics and Data Science, Yale School of Medicine, Yale

University, New Haven, USA
[19] Ophthalmology and Visual Sciences (Eye ACP), Duke-NUS Medical School, Singapore

**Corresponding author:**

Prof Pearse A. Keane
Institute of Ophthalmology, University College London, London, UK.
Email: p.keane@ucl.ac.uk

Dr Yih Chung Tham
Yong Loo Lin School of Medicine, National University of Singapore.
Tel: +65 65767298, Fax: +65 6225 2568; Email: thamyc@nus.edu.sg

**Financial Support:**

Prof Pearse A. Keane is supported by the UK Research & Innovation Future Leaders Fellowship (MR/T019050/1), the Moorfields Eye Charity with The Rubin Foundation Charitable Trust (GR001753), and an Alcon Research Institute Senior Investigator Award. Dr. Yih Chung Tham is supported by the National Medical Research Council of Singapore (NMRC/MOH/ HCSAINV21nov-0001 & HCSAINV25jan-003). The funder had no role in study design, data collection, data analysis, data interpretation, writing of the report, or the decision to submit for publication.

**Conflict of Interest:**

Dr. Pearse A. Keane has acted as a consultant for Retina Consultants of America, Topcon, Roche, Boehringer-Ingleheim, and Bitfount and is an equity owner in Big Picture Medical. He has received speaker fees from Zeiss, Novartis, Gyroscope, Boehringer-Ingleheim, Apellis, Roche, Abbvie, Topcon, and Hakim Group. He has received travel support from Bayer, Topcon, and Roche. He has attended advisory boards for Topcon, Bayer, Boehringer-Ingleheim, RetinAI, and Novartis.

**Ethics statement:**

All included original, non open-source studies adhered to the tenets of the Declaration of Helsinki and had respective local ethical committee approval. We obtained permission from the principal investigator of each study to use the data.

**Abstract**

**Purpose:** DINOv2 is a natural image-based foundation model (FM), pre-trained exclusively on 142 million natural images from the LVD-142M dataset. In contrast, RETFound is a retina-specific FM, pretrained on ~3 million images, including natural images, colour fundus photos (CFPs), and optical coherence tomography (OCT) images (~1 million each). Despite DINOv2's massive pretraining dataset, its application in ophthalmology and relative performance to domain-specific FMs remain understudied. To address this gap, we conducted a head-to-head comparative evaluation between DINOv2 and RETFound models across a range of downstream ocular and systemic disease tasks.

**Design:** Retrospective, head-to-head evaluation.

**Subjects, Participants and/or controls:** Ocular disease detection tasks included diabetic retinopathy (DR), glaucoma, and multi-class eye diseases, while systemic disease incidence prediction focused on the 3-year incidence of heart failure, myocardial infarction, and ischemic stroke. Eight open-source datasets (APTOS-2019, IDRID, MESSIDOR2 for DR; PAPILA, Glaucoma Fundus for glaucoma; JSIEC, Retina, OCTID for multi-class eye diseases) and the Moorfields AlzEye dataset (for systemic diseases), were used for fine-tuning and internal testing. External test sets included the same open-source datasets (cross-dataset validation) and the UK Biobank (for systemic diseases).

**Methods:** We replicated the fine-tuning methodology from the original RETFound study on three DINOv2 models (Large, Base, Small). All models were fine-tuned on the respective datasets and evaluated through internal and external testing.

**Main outcome measures:** Area under the receiver operating characteristics curve (AUROC) and two-sided t-tests were used to compare models' performances.

**Results**: For ocular disease detection, DINOv2 models generally outperformed RETFound. For DR, DINOv2-Large achieved AUCs of 0.850–0.952, exceeding RETFound's 0.823–0.944 (all $p \leq 0.007$). For multi-class eye diseases, DINOv2-Large (AUC=0.892, Retina dataset) surpassed RETFound (AUC=0.846, $p<0.001$). For glaucoma, DINOv2-Base (AUC=0.958, Glaucoma Fundus) outperformed RETFound (AUC=0.940, $p<0.001$). Conversely, for systemic disease incidence prediction, RETFound achieved superior AUCs of 0.796 (heart failure), 0.732 (myocardial infarction), and 0.754 (ischemic stroke), outperforming DINOv2's best models' AUC (0.663–0.771, all $p<0.001$). This trend persisted in external validation.

**Conclusion:** Our findings reveal the merits of DINOv2 in ocular disease detection tasks, while RETFound demonstrates an edge in systemic disease incidence prediction. These findings showcase the distinct scenarios where general-purpose and domain-specific FMs excel, highlighting the importance of aligning FM selection with task-specific requirements to optimise clinical performance.

## Introduction

The advent of foundation models (FMs) have ushered in a paradigm shift in medical deep learning (DL) approaches[1][2][3][4]. Driven by the availability of large-scale high-quality unstructured datasets and advanced self-supervised learning methods, FMs excel in performing multiple tasks with high adaptability and generalizability to new, unseen tasks, often requiring minimal fine-tuning [5][6]. In ophthalmology, the introduction of RETFound, the first retina-specific FM developed by Zhou et al. [7] in 2023, led to a significant transformation in DL approaches for medical image analysis. Using a Vision Transformer (ViT) [8][9] backbone and a Masked Autoencoder (MAE)-based [10] self-supervised learning approach, RETFound was pre-trained on a diverse dataset of 1.4 million natural images (ImageNet-1K), ~900,000 color fundus photographs (CFPs), and ~730,000 OCT scans. This pretraining approach enabled RETFound to effectively extract and learn retinal features and representations, enhancing diagnostic performance, data efficiency, and generalizability across ophthalmic and systemic clinical tasks [11][12][13].

In parallel, DINOv2 is a general-purpose vision FM developed by Meta AI, pre-trained on 142 million natural images from the LVD-142M dataset [14]. With its sheer scale and versatile feature representation capabilities, DINOv2 has demonstrated strong adaptability across some medical domains, such as pathology [15][16] and radiology[17]. However, the applicability of DINOv2 to ophthalmic tasks remains largely unexplored, and a direct comparison with RETFound in fine-tuning for retinal image analysis has not yet been performed.

To address these gaps, we conducted a head-to-head performance comparison between three DINOv2 models (large, base, small) and RETFound across a range of ocular disease detection and systemic disease incidence prediction tasks. For a fair and standardized comparison, we replicated the RETFound study's original fine-tuning methodology on all three DINOv2 models (**Figure 1**). The models' comparative performances were evaluated across eight open-source datasets for ocular disease detection tasks including diabetic retinopathy (APTOS-2019, IDRID, MESSIDOR2), glaucoma (PAPILA, Glaucoma Fundus), and multi-class eye diseases (JSIEC,

Retina, OCTID). For systemic disease incidence prediction tasks (3-year incidence of heart failure, myocardial infarction, ischemic stroke), we fine-tuned all models on the Moorfields AlzEye dataset, and externally tested on the UK Biobank dataset. Model performance was primarily measured using the area under receiver operator characteristics curve (AUROC), and the significant differences in AUROC values between models were determined through two-sided t-tests with Bonferroni correction. This comprehensive evaluation provides critical insights into the respective strengths and optimal applications of general-purpose and retina-specific FMs, contributing new understanding to the utilization of FM in ophthalmology and beyond.

## Methods

### Details of pretraining datasets for DINOv2 models

All three model versions of DINOv2 - Small, Base, and Large - were pretrained on the LVD-142M dataset, which is a large-scale unannotated dataset with a total of 142,109,386 natural images [14]. This massive dataset was curated from multiple open-source natural image databases, including the ImageNet-22k (56,788,344 images, 40.0%), the ImageNet-1K[18] (40,997,344 images, 28.8%), the Google Landmarks v2 (6,321,880 images, 4.4%), and Mapillary SLS (1,434,262 images, 1.0%) datasets, and the remaining 36,567,556 images (25.8%) obtained through retrieval and augmentation of multiple smaller datasets. To ensure data quality and diversity, researchers employed self-supervised image retrieval and clustering methods to filter and rebalance the curated dataset, which provided rich visual feature learning resources for pretraining of the DINOv2 models.

### Task-specific datasets for ocular disease detection.

We evaluated model performance on multiple publicly available ophthalmological datasets covering various eye diseases. For DR detection, we used three datasets, the IDRiD (India)[19][20], MESSIDOR-2 (France)[21][22][23], and APTOS-2019 (India) datasets[24]. The DR grading for these datasets were based on the International Clinical Diabetic Retinopathy Severity Scale, which indicated five stages of severity from no DR to proliferative DR. For detection of glaucoma, we used the PAPILA (Spain)[25] and Glaucoma Fundus (Korea) datasets[26]. Both datasets have glaucoma labels of three classes, non-glaucoma, early glaucoma (suspected glaucoma), and late-stage glaucoma. For detection of multi-category eye diseases, we used two CFP datasets including the JSIEC dataset (China)[27] which comprises 1,000 images covering 39 classes of common referable fundus diseases and conditions, and the Retina dataset [28], comprising 601 images with normal, glaucoma, cataract, and retinal diseases annotations. An additional OCT dataset, the OCTID dataset (India)[29], containing 470 OCT scans with labels including normal, macular hole, age-related macular degeneration (AMD), central serous retinopathy, and DR was also used. For more detailed information about the datasets used, please refer to Table S1.

**Task-specific datasets for incidence prediction of systemic diseases.**

For incidence prediction of systemic diseases, we focused on three cardiovascular disease outcomes: heart failure, myocardial infarction, and ischemic stroke. We used retinal images to predict the risk of these three cardiovascular diseases occurring within the next 3 years. All models were fine-tuned on a curated private dataset from the Moorfields AlzEye[30] study (MEH-AlzEye) and internally evaluated on a held-out test dataset. The MEH-AlzEye study is a large-scale retrospective cohort study that linked the ophthalmological data of 353,157 patients seen at Moorfields Eye Hospital between 2008 and 2018 with systemic health data across England. These systemic health data primarily come from the Hospital Episode Statistics (HES) data that documented hospital records of patients receiving inpatient care. Additionally, we leveraged the UK Biobank for external evaluation of three systemic diseases, a large-scale biomedical database that includes data from 502,665 participants aged 40 to 69 years who are registered with the UK National Health Service. Among them, 82,885 participants underwent retinal imaging and OCT examination, and a total of 171,500 retinal images were obtained. For consistency, we selected one left-eye retinal image per patient visit for model training, to avoid bias caused by inconsistent individual visits. We divided patient data into training, validation, and test sets for model training, tuning, and evaluation. This approach allowed us to assess the potential of retinal images in predicting these important systemic diseases, providing new insights for early intervention and personalized medical strategies. For more detailed information about the datasets, please refer to

Table S2.

**DINOv2 model architecture and implementation**

DINOv2 employs Vision Transformer (ViT) as its backbone network with three different model size scales: Small, Base, and Large. DINOv2-Small (21M parameters) is based on the ViT-S/14 architecture, with an embedding dimension of 384, with 6 attention heads, and 12 Transformer blocks. DINOv2-Base (86M parameters) utilized the ViT-B/14 architecture, featuring an embedding dimension of 768, with 12 attention heads, and 12 Transformer blocks. DINOv2-Large (303 M parameters) was built on the ViT-L/14 architecture, with the largest embedding dimension of 1024, along with 16 attention heads, and 24 Transformer blocks. All three model versions use a patch size of 14×14. During training, DINOv2 employs the AdamW optimizer with an initial learning rate of 1e-3, a batch size of 2048, and a total of 625k iterations. To enhance performance, LayerScale technology was introduced with an initial value of 1e-5. Weight decay follows a cosine schedule ranging from 0.04 to 0.2. Learning rate warm-up lasts for 100k iterations, while the teacher network momentum also adopts a cosine schedule, gradually increasing from 0.994 to 1. The models were primarily trained in float16 precision, but the gradients for DINO heads were computed in float32 precision to ensure accuracy. The feed-forward network layers used a multilayer perceptron (MLP) structure with a dropout rate of 0 to maximize information transfer. These models were distilled from a larger ViT-G/14 model, effectively learning and extracting powerful visual feature representations while maintaining training efficiency.

**Details on task-specific fine-tuning of the DINOv2 models.**

When adapting to downstream tasks, we used all three versions of the DINOv2 model (Large, Base, or Small) to generate high-level features from the retinal images, and added an MLP block to each model. The MLP block takes these high-level features as input and generates probability scores (ranging from 0 to 1) as outputs in each ocular disease or systemic disease incidence prediction task. The number of neurons in the last layer of the MLP block is determined by the number of categories in each task. To enhance the models' generalization, we also applied various data augmentation techniques, including random cropping, horizontal flipping, vertical flipping, rotation, and adjustments to brightness and contrast. The batch size was set to 32 and training was conducted over 100 epochs. The first 10 epochs was a learning rate warm-up phase, increasing from 0 to $5\times10^{-3}$. For the subsequent 90 epochs, a cosine annealing scheduling was applied, gradually decreasing the learning rate from $5\times10^{-3}$ to $1\times10^{-6}$. We used the AdamW optimizer with weight decay set to 0.05. After each training epoch, the models were evaluated on the validation set and the model iteration with the highest AUROC on the validation set was used for all downstream evaluation. These model weights were then implemented for independent internal and external evaluations. For RETFound, we utilized the publicly available pre-trained weights of the model for each specific task.

For the cross-dataset evaluation of DR, glaucoma detection, and systemic disease incidence prediction, we employed the same fine-tuning strategy but trained and tested the models on different dataset combinations. Specifically, the datasets used for DR include APTOS-2019, MESSIDOR-2, and IDRiD; for glaucoma, Glaucoma_fundus and PAPILA; and for systemic diseases, Moorfields AlzEye and UK Biobank.

In the label efficiency analyses, we applied varying percentages of training data (from 10% to 100%) and finetuned the models for each ocular disease detection and systemic disease incidence prediction tasks. This label efficiency analysis was repeated across all datasets (eight open-source ocular disease datasets and the Moorfields AlzEye dataset), and all models were internally validated on held-out test sets. This enabled a comprehensive analysis of the models' performance in scenarios with limited labeled data for task-specific finetuning.

**Performance evaluation and statistical analysis**

For all tasks, we used the area under receiver operating characteristics curve (AUROC) to evaluate the models' diagnostic performance. For multi-class detection tasks, including detection of DR (5 classes), glaucoma (3 classes), multi-category eye diseases and systemic diseases, we calculated the AUROC for each class and then averaged across the number of classes to obtain the overall AUROC. We calculated the mean and standard deviation of the AUROC by randomly sampling 20% of the data in each iteration across 100 bootstrap replicates. The standard error was obtained by dividing the standard deviation by $\sqrt{100}$, and the 95% confidence interval was computed as the mean ± 1.96 × standard error. Normality was evaluated using the Shapiro-Wilk test applied to the AUROC derived from 100 bootstrapped replicate samples, with the majority demonstrating a normal distribution. To establish statistical differences in the AUROC performance between the DINOv2 and RETFound models, we computed the p-value (P) using the two-sided t-test with Bonferroni correction. Statistical significance was defined as P less than 0.05/3 (P=0.017), accounting for multiple pairwise comparisons between RETFound and the DINOv2 models. To evaluate the model calibration performance, we calculated the Expected Calibration Error (ECE) and plotted the reliability diagrams to ensure that the model's predictive accuracy matches its confidence level. For external evaluations of ocular and systemic diseases, we also reported metrics such as Kappa, F1-score, and accuracy.

## Results

The demographic characteristics and details of study datasets utilized are detailed in **Table S1 and S2**.

### Comparative Performance of Models in Ocular Disease Detection Tasks:

DINOv2 fine-tuned models generally outperformed RETFound across ocular disease detection tasks in internal and external testing (**Figure 2, Figure S3-S6, and Table S3-S5**).

For diabetic retinopathy (DR) detection task, DINOv2-Large achieve higher AUROC values across the datasets. For example, after fine-tuning and internally validating using the APTOS2019 dataset, DINOv2-Large achieved an AUROC of 0.952 (95% CI: 0.950–0.954), surpassing RETFound (AUROC: 0.944, 95% CI: 0.941–0.946, all P<0.001, **Figure 2 and Figure S3, Table S3**). Consistently, in the MESSIDOR-2 dataset, DINOv2-Large achieved an AUROC of 0.906 (95% CI: 0.902–0.910), higher than RETFound (AUROC: 0.883, 95% CI: 0.878–0.889, P < 0.001). Similarly, in the IDRID dataset, DINOv2-Large obtained an AUROC of 0.850 (95% CI: 0.836–0.865), outperforming RETFound (AUROC: 0.823, 95% CI: 0.809–0.836, P = 0.007).

**Figure S4** shows the confusion matrices of all models in the internal test for each ocular disease tasks, at a threshold of 0.5. Similarly, in external validations, where cross-evaluation was performed by fine-tuning models on one dataset and testing them on the other two (repeating this permutation across all three datasets), DINOv2 models generally performed better than RETFound (Figure S5 **and Table S4)**. For instance, when fine-tuned on APTOS2019 and externally tested MESSIDOR2, DINOv2-Large achieved the highest AUROCs of 0.817, outperforming RETFound (AUROCs: 0.725 in MESSIDOR2, **Figure S5a**, all p<0.001). When fine-tuned on IDRID and tested on APTOS2019, DINOv2-Large (AUROC:0.783) and DINOv2-Small (AUROC:0.808) outperformed RETFound (AUROC:0.752, all P<0.001). Evaluations based on accuracy, Kappa values, and F1-score, consistently showed that DINOv2 models outperformed RETFound in most scenarios (**Figure S5b-d**).

For the glaucoma detection task, when fine-tuned and internally validated on the Glaucoma_fundus dataset, DINOv2 models achieved higher AUROC scores than RETFound (**Figure 2**). Specifically, DINOv2-Base and -Small achieved AUCs of 0.958 (95% CI: 0.955–0.961), outperforming RETFound (AUROC: 0.940, 95% CI: 0.936–0.945, all P<0.001, **Figure 2 and Table S3**). Similarly, in external validation, when fine-tuned on the Glaucoma_fundus dataset and tested on the PAPILA dataset, DINOv2-Large achieved the highest performance with an AUROC of 0.625, surpassing RETFound's 0.558 (P<0.001, **Figure S6a**, **Table S5**). Evaluations based on accuracy, Kappa values, and F1-score, consistently largely showed that DINOv2 models outperformed RETFound in most scenarios (**Figure S6b-d**).

For the multi-class eye disease classification tasks, the models were finetuned and validated internally on datasets containing CFPs (the JSIEC and Retina datasets) and OCT scans (**Figure 2**). In the JSIEC CFP dataset, all models were comparable and achieved AUROC scores close to 1. In the Retina CFP dataset, the DINOv2 models achieved AUROC scores ranging from 0.874 (DINOv2-Base: 95% CI: 0.864–0.883) to 0.892 (DINOv2-Large: 95% CI: 0.883–0.902), surpassing RETFound's performance (AUROC: 0.846, 95% CI: 0.836–0.856, all P≤0.001). For the OCTID dataset with OCT scans, all models showed comparable performance.

**Comparative Performance of Models in Predicting 3-year incidence of Systemic Diseases:**

We further evaluated the comparative performance of DINOv2 and RETFound models in predicting three cardiovascular-related systemic diseases - heart failure, myocardial infarction, and ischemic stroke. Across all three systemic disease outcomes, the models were fine-tuned and tested internally on the Moorfields AlzEye study dataset, and externally validated using the UK Biobank dataset **(Figure 7 & Figure S8-S9 & Table S6**).

For heart failure prediction, in internal testing, RETFound significantly outperformed all DINOv2 variants, achieving an AUROC of 0.796 (95% CI: 0.767-0.827), compared to DINOv2-Base's 0.771 (95% CI: 0.738-0.807), the best-performing DINOv2 model ($P < 0.001$, **Figure 7 and Figure S8).** For myocardial infarction prediction, RETFound achieved the highest AUROC of 0.732 (95% CI: 0.663-0.805), significantly outperforming the best DINOv2 model, DINOv2-Large (AUROC: 0.711, 95% CI: 0.638-0.784, $P < 0.001$). For ischemic stroke prediction, RETFound demonstrated consistent superiority with an AUROC of 0.754 (95% CI: 0.684-0.838), surpassing the best DINOv2 model, DINOv2-base (AUROC: 0.714, 95% CI: 0.633-0.789, $P < 0.001$).

In the external validation using the UK Biobank dataset (**Figure S9 and Table S6**), RETFound consistently outperformed all DINOv2 models across the three systemic disease incidence prediction tasks. For heart failure prediction, RETFound achieved the highest AUROC (0.674), surpassing all DINOv2 models (AUROC: 0.615-0.623, all $P<0.001$). Similarly, for myocardial infarction prediction, RETFound maintained its advantage with an AUROC of 0.594, outperforming all DINOv2 models (AUROC: 0.523-0.559, all<0.001). In ischemic stroke prediction, RETFound achieved an AUROC of 0.586, exceeding DINOv2-Base (0.556) and DINOv2-Small (0.519) (all $P \leq 0.008$). Secondary metric, including accuracy, further reinforced RETFound's superiority, with higher values compared to all DINOv2 models across all three systemic outcomes (**Figure S9**).

**Comparative Performance in Label Efficiency:**

To investigate the impact of training data size on FMs performance, we conducted an in-depth analysis using varying proportions of training/ fine-tuning data (10%, 20%, 50%, 90%, and 100%) across different ocular disease detection and systemic disease incidence prediction tasks (**Figure 10 and Table S7-S8**).

For DR detection task, in the APTOS2019 dataset, the DINOv2 models and RETFound achieved comparable AUROC performance, with minimal variation across different proportions of training data (**Figure 10 and Table S7)**. However, in the MESSIDOR2 and IDRID datasets, the DINOv2 models consistently outperformed RETFound, achieving higher AUROCs as the training data proportion decreased. Furthermore, in the APTOS2019 and MESSIDOR2 datasets, the performances of DINOV2 models remained relatively stable (i.e. no significant drop) despite decrease in training data size. In the multi-class eye disease detection tasks (e.g., JSIEC, Retina, OCTID), DINOv2 models consistently exhibited higher AUROCs than RETFound across nearly all training data proportions. Notably, DINOv2 models required fewer training samples to achieve superior performance (particularly in the OCTID dataset), demonstrating better label efficiency in these multi-class tasks. Nevertheless, in glaucoma detection task, all models largely showed similar performances across the scenarios.

On the other hand, in systemic disease incidence prediction tasks, RETFound exhibited a distinct advantage where it consistently achieved higher AUROC values across all training data proportions (**Figure 10 and Table S8**). Notably, as training data decreased, RETFound maintained relatively stable or similar performances across the three systemic prediction tasks, reflecting its robustness. At lower training proportions (e.g., 10%–50%), DINOv2 models, particularly DINOv2-Small, exhibited more variability and consistently underperformed compared to RETFound.).

**Comparison of Calibration Performance between DINOv2 and RETFound Models**

We compared the calibration performance of DINOv2 models (Large, Base, and Small) with RETFound across ocular and systemic disease tasks using corresponding datasets. Overall, we observed that DINOv2 models generally demonstrated better calibration for ocular diseases, while RETFound exhibited superior calibration for systemic disease incidence prediction.

For DR detection task, in the APTOS2019 dataset, it was revealed that DINOv2-Base and -Small were better calibrated than RETFound, as indicated by their calibration curves being closer to the perfectly calibrated line and their lower expected calibration error (ECE) values (**Figure 11, Figure S12-S13**). In the IDRID dataset, DINOv2-Small demonstrated the best calibration performance with its curve closely following the diagonal line. Similarly in glaucoma detection, DINOv2-Base (ECE: 0.015) performed best, especially in high-probability regions in the Glaucoma_fundus dataset, while DINOv2-Large (ECE:0.040) achieved superior overall calibration on the PAPILA dataset, outperforming RETFound (ECE=0.065). On the other hand, for datasets covering multiple retinal diseases, such as Retina, DINOv2-small (ECE:0.013) had more superior calibration than RETFound (ECE:0.027). In the JSIEC and OCTID datasets, all models exhibited fluctuations and had similar calibration.

Across the tasks of predicting heart failure and myocardial infarction (**Figure 11**), RETFound demonstrates stable calibration performance. Its calibration curves were closer to the perfectly calibrated line, and it achieved ECE values mostly lower than DINOv2 models' values. Nevertheless, when predicting ischemic stroke, RETFound exhibited poorer calibration compared to its performance in other systemic disease tasks, with an ECE of 0.212. By contrast, DINOv2-Base achieved the best calibration among all models with an ECE of 0.024, followed by DINOv2-Small (ECE = 0.125). DINOv2-Large, however, displayed the highest calibration error among all models, with an ECE of 0.312, indicating substantial miscalibration in its probability estimates for ischemic stroke prediction.

## Discussion

In this study, we evaluated the comparative performance and label efficiency between fine-tuning on DINOv2, a general-purpose vision FM, with RETFound, a retina-specific FM, across a range of ocular disease detection and systemic disease incidence prediction tasks. We ensured a standardized and robust comparison with RETFound by replicating RETFound study's original fine-tuning methodology on all three DINOv2 models (large, base, small). This is a rare feat and has not been conducted prior. Overall, in ocular disease detection tasks, we observed that DINOv2-large outperformed RETFound and with better label efficiency. Additionally, DINOv2-base and -small models generally required three times less computing resources compared to RETFound. Conversely, in systemic disease incidence prediction tasks, RETFound generally achieved superior performance and better label efficiency than all the DINOv2 models. Ultimately, this comparative performance elucidates the complementary strengths of both FMs' approaches: DINOv2 leverages the scale and diversity of natural image pretraining, while RETFound incorporates domain-specific retinal 'subtle knowledge'. These findings underscore the need to align model choice, fine-tuning strategy, label efficiency and computational scalability with the nature and complexity of the clinical task.

In ocular disease detection tasks, DINOv2 generally outperformed RETFound, when fine-tuned on the complete dataset, highlighting the transferability of DINOv2 to these tasks. Additionally, DINOv2 demonstrated exceptional label efficiency, maintaining robust AUROC performance even with limited training data (**Figure 2 and 10, Table S7**). Ocular disease detection tasks often depend on identifying well-defined lesions (e.g., microaneurysms, hemorrhages) with distinct visual patterns, which are features that general-purpose models pretrained on natural images can effectively capture. While it may seem counterintuitive that RETFound, despite having pretrained on retinal images, achieved comparable performance to DINOv2 models in ocular disease detection tasks, it is likely due to the differences in pretraining strategy and data scale. In DINOv2's context, the large corpus of 142 million natural images in the LVD-142M dataset enabled DINOv2 to learn diverse visual patterns and contexts, enhancing its feature extraction capabilities and thereby establishing a stronger foundation for various downstream visual tasks. This could have contributed to DINOv2's exceptional transfer learning capabilities, a finding consistent with prior research on AMD and DR detection[31][32], as well as across other medical imaging modalities, including X-ray, MRI and ultrasound[33][34]. Furthermore, DINOv2 employs a self-distillation process where knowledge transfer occurs between the teacher and student network, enabling the model to efficiently learn refined visual representations. The broad knowledge of visual representations acquired during its larger pretraining could have enhanced DINOv2's generalizability to ocular disease detection tasks. These findings suggest that DINOv2 could serve as a competitive alternative to RETFound particularly in low-data resource settings for specific ocular diseases detection tasks. More broadly, they underscore the potential of general-purpose FMs as practical and scalable bases for developing downstream disease-specific detection models, especially in settings where annotated medical data are limited.

Interestingly, while DINOv2 demonstrated superior performance over RETFound in the ocular disease diagnosis task, this trend was not fully replicated in all external validations (**Figure S4 and S5**). For instance, the performance degradation observed in glaucoma detection, most notably the drop from internal AUC = 0.958 on Glaucoma_Fundus to external AUC = 0.625 on PAPILA. Such performance degradation is likely attributable to substantial differences in image characteristics across the two datasets, including original resolution size and field of view (**Figure S14**). Moreover, the clinical definitions and diagnostic criteria for glaucoma are not entirely the same between PAPILA and Glaucoma_Fundus, contributing further heterogeneity in label consistency and ground truth interpretation. Taken together, these domain-level variations indicate that both DINOv2 and RETFound are also susceptible to distribution shifts like conventional deep learning models, when deployed across heterogeneous populations, imaging modalities, and clinical definitions, thus emphasizing the need for more robust cross-domain benchmarking in future studies.

Additionally, in terms of computational resource consumption (**Table S9**), RETFound required more GPU RAM during the fine-tuning phase but had a lower inference speed (i.e. took more time to process the same number of images) as compared to the DINOv2-small and -base. DINOv2-small (22M parameters) and DINOv2-base (86M parameters) are significantly smaller models compared to RETFound and DINOv2-large (both models have 303M parameters). As such, when tested on 100 retinal images (with 224×224 resolution and processed on a single A100 GPU), DINOv2-small (0.48 seconds) and DINOv2-Base (0.64 seconds), both had lower inference time compared to RETFound (1.48 seconds) and DINOv2-Large (1.70 seconds). This underscores the potential of smaller and more computational resource efficient general-purpose vision FMs like DINOv2-small and base as a competitive alternative to RETFound for ocular disease detection tasks under resource-scarce settings. More importantly, it highlights the need to consider the trade-off between model performance and computational resource requirements depending on unique deployment sites.

On the other hand, in systemic disease incidence prediction tasks, RETFound consistently outperformed the DINOv2 models across all varying proportions of fine-tuning data used, indicating RETFound's superior label efficiency and its ability to detect subtle retinal changes associated with systemic diseases. (**Figure 7 and 10, Table S8**). Such tasks involve detecting more subtle and complex markers, such as vascular structures or hidden patterns embedded within the retina (e.g., vascular caliber, vessel tortuosity, and branching patterns). These context-rich features may not be adequately captured by models pretrained on natural images or simpler model architectures. As such, RETFound's superior performance is likely attributed to its pretraining strategy, which harnessed domain-specific knowledge from 900,000 unlabeled retinal images using the MAE-supervised learning approach [7]. The MAE approach involves having the model reconstruct highly masked input images, forcing the model to infer missing information from the masked patches of the images during training. Through this iterative training process, the model learns to extract subtle, fine-grained vasculature patterns and features from the masked retinal images, thereby potentially contributing to RETFound's retina-specific domain knowledge. For oculomics tasks, RETFound superior label efficiency demonstrated its potential as an effective pre-trained model, especially given well-annotated data still remain scarce and expensive to curate. Overall, these findings suggest the advantages of incorporating domain-specific data in the pretraining process of FMs, which may yield significant performance gains for complex downstream clinical applications.

To illustrate the practical benefits of pretraining on FMs, using the same fine-tuning datasets as the main analyses, we observed that models loading pre-trained weights (DINOv2 and RETFound) consistently outperformed models trained from scratch, across all ocular and systemic disease tasks (**Figure S15 and Figure S16**). These observations reaffirm that large-scale pretraining, irrespective whether based on natural images or retina-specific corpora, is foundational for achieving strong and reliable downstream performance. Nonetheless, while FMs represent an important step forward and are closing the translational gap, we believe prospective implementation studies remain essential to assess clinical utility in actual settings. These include understanding clinician acceptance, patient trust, cost-effectiveness, and health system integration which are critical elements that will ultimately determine whether these models are truly ready for clinical adoption.

Our study's key strength lies in replicating the intricate methodology of the original RETFound experiments, ensuring consistent and robust head-to-head evaluation between DINOv2 and RETFound across diverse ocular disease detection and systemic disease incidence prediction tasks. We fine-tuned the DINOv2 models using multiple open-source datasets, applying the same fine-tuning and test data split as applied in RETFound's task-specific adaptions. We intentionally focused on predicting the 3-year incidence of systemic conditions using datasets from Moorfields Eye Hospital (AlzEye) and the UK Biobank. These tasks are more representative of clinical prognostication than simpler cross-sectional classification tasks. Importantly, these two datasets were also used in the original RETFound study, and we chose to retain them to ensure a fair, objective, and reproducible head-to-head comparison between RETFound and DINOv2. AlzEye and UK Biobank remain among the few large-scale, ICD-validated resources linking retinal imaging to systemic health outcomes and represent some of the most robust datasets currently available for this purpose. To our knowledge, such a benchmarking evaluation of a general-purpose vision FM and RETFound has never been conducted to date. This open approach provides credibility to the repeatability of our study findings, and would help to facilitate comparable benchmarking evaluations for future FMs.

Nevertheless, this study has a few limitations. In this study, the range of detection tasks included was limited by the availability of annotated open-source CFP and OCT datasets with specific disease labels. Therefore, the comparative performance of DINOv2 and RETFound models was not demonstrated on other modalities such as ultra-wide field retinal imaging. Furthermore, even though we evaluated the models' performance in detecting multiple ocular diseases and predicting multiple systemic diseases, other rare ocular (e.g. Retinitis Pigmentosa) and systemic diseases were not included in this study. Future work should include benchmarking of the performance of FMs on other conditions such as dementia and coronary artery diseases. Additionally, the comparison between DINOv2 and RETFound likely reflects two major underlying factors: first, difference in the scale and content of the pretraining datasets (natural vs retinal images), and second, differences in pretraining strategy (contrastive learning based versus generative learning based). It will be interesting to disentangle the impact of the two factors and fully understand the technical advantages of DINOV2 versus RETFound, however, this has been hampered by limited access to the LVD-142M dataset (i.e. this dataset was carefully curated through complicated engineering pipeline[14] and is currently not publicly available). Building on our findings, to deepen insights on the clinical applicability of these FMs, it will be critical to evaluate their comparative performance using datasets that are representative of actual clinical setting. Even though both DINOv2 and RETFound demonstrated strong performance in ocular disease detection, the observed AUC differences in diabetic retinopathy, are numerically modest. While some of these differences may reach statistical significance, they may not always reflect meaningful clinical superiority, particularly when confidence intervals overlap. In such high-performing models (AUC > 0.90), incremental differences should also be interpreted with caution and in the context of downstream use, disease prevalence, and operational workflows. For systemic disease risk prediction, however, even modest AUROCs (e.g., 0.70–0.80) can still have significant value in applications such as population screening, early risk stratification, or referral triage especially when used alongside other clinical features or biomarkers.

Future work would involve the refinement of domain adaptation strategies and harmonized, multi-institutional pretraining efforts, to improve the generalizability of these FMs for diverse clinical task scenarios across the medical domain and wider healthcare contexts. Globally, this will require collaborative efforts from researchers worldwide to curate diverse datasets, adopt standardized reporting practices to ensure reproducibility and comparability, and engage stakeholders in collectively democratizing access to datasets and advanced FMs[35].

## Conclusion

This study provides a comprehensive evaluation of general-purpose vision FMs (DINOv2 models) and a retina-specific FM (RETFound), highlighting DINOv2's optimal adaptability for ocular disease detection tasks and RETFound's superior performance in systemic disease incidence prediction. These findings provide new insights into the optimal utilisation of general-purpose vision FM and retina-specific FM, highlighting the importance of aligning FM selection with task-specific requirements and resource considerations to achieve optimal clinical performance.

## Data availability statement

The AlzEye dataset is governed by contractual restrictions between NHS Digital, Moorfields Eye Hospital, and University College London, limiting access to the AlzEye research team. While collaborations are welcomed, individual-level systemic health data can only be analyzed by AlzEye researchers. Detailed information about data curation can be found in the RETFound paper and its supplementary materials. For more information about the dataset and potential collaborations, please visit https://readingcentre.org/studies/artificial_intelligence/alzeye or refer to the comprehensive documentation in Nature (https://www.nature.com/articles/s41586-023-06555-x#Sec10) and its supplementary figures (https://static-content.springer.com/esm/art%3A10.1038%2Fs41586-023-06555-x/MediaObjects/41586_2023_6555_MOESM1_ESM.pdf). UK Biobank data can be accessed at https://www.ukbiobank.ac.uk/.

Publicly available datasets for ocular disease experiments include:

IDRID: https://ieee-dataport.org/open-access/indian-diabetic-retinopathy-image-dataset-idrid;

MESSIDOR-2: https://www.adcis.net/en/third-party/messidor2/;

APTOS-2019: https://www.kaggle.com/competitions/aptos2019-blindness-detection/data;

PAPILA: https://figshare.com/articles/dataset/PAPILA/14798004/1;

Glaucoma Fundus:
https://dataverse.harvard.edu/dataset.xhtml?persistentId=doi:10.7910/DVN/1YRRAC;

JSIEC: https://zenodo.org/record/3477553;

Retina: https://www.kaggle.com/datasets/jr2ngb/cataractdataset;
OCTID: https://borealisdata.ca/dataverse/OCTID.

**Code availability statement**

The training, fine-tuning, and evaluation code for RETFound, provided by Y.Z. and built on PyTorch, is available at https://github.com/rmaphoh/RETFound_MAE. The corresponding code for DINOv2 is available at https://github.com/HouQingshan/DINOv2.

**Figure legend**

**Figure 1. Overview of the development and evaluation of the DINOv2 and RETFound models.** Stage 1 illustrates the pre-training phase of DINOv2 and RETFound models through self-supervised learning on respective natural and ophthalmic image datasets. Stage 2 covers the supervised fine-tuning of the pre-trained models for adaptation to specific downstream ocular disease diagnosis and systemic disease incidence prediction tasks.

**Figure 2. AUROC performance of DINOv2 and RETFound for ocular disease detection tasks in the internal tests.** We fine-tuned DINOv2-Large, DINOv2-Base, DINOv2-Small, and RETFound models separately to adapt to each dataset, and internally evaluated them on hold-out test data for ocular disease diagnostic tasks, including diabetic retinopathy, glaucoma, multi-category eye diseases. For each task, statistical significance of AUROC differences was evaluated using 100 bootstrap replicates, randomly sampling 20% of the data. Error bars show 95% confidence intervals, and bar centers represent the mean AUROC values.

**Figure 7. AUROC performance of DINOv2 and RETFound for systemic disease incidence prediction tasks (3-year incidence) in the internal tests**. The models were fine-tuned using MEH-AlzEye and internally evaluated on hold-out test data. For each task, statistical significance of AUROC differences was evaluated using 100 bootstrap replicates, randomly sampling 20% of the data. Error bars show 95% confidence intervals, and bar centers represent the mean AUROC values.

**Figure 10. Label efficiency of DINOv2 and RETFound for ocular disease detection and systemic disease incidence prediction tasks.** The models were fine-tuned with varying proportions of task-specific training data (from 10% to 100%) and internally evaluated on hold-out test data. Such analysis is key in optimizing the amount of data required to reach optimal performance levels.

**Figure 11. Reliability diagrams and expected calibration error (ECE) for ocular disease**

**detection and systemic disease incidence prediction tasks**. Calibration plots (10 bins, full test set) are shown for DINOv2 (Large, Base, Small) and RETFound models, with the dashed diagonal line representing perfect calibration. The ECE values quantify model miscalibration, where lower values indicate better alignment. These plots highlight performance differences among models across eight open-source ocular disease datasets and the Moorfields AlzEye dataset for systemic diseases. The DINOv2 models generally performed comparably or better than RETFound with lower ECE values in ocular disease tasks. While RETFound largely achieved lower ECE values than the DINOv2 models in systemic disease tasks, with an exception when compared to DINOv2-Base in predicting ischemic stroke.